\documentclass[preprint,amsmath,amssymb]{revtex4}
\usepackage{graphicx}
\usepackage{bm}

\newcommand{\E}[1]{Eq.~(\ref{#1})}
\newcommand{\F}[1]{Fig.~\ref{fig:#1}}
\begin{document}
\title{Swarm-Oscillators}
\author{Dan Tanaka}
 \email{dan@ton.scphys.kyoto-u.ac.jp}
 \affiliation{%
Department of Human and Artificial Intelligent Systems (HART),
Graduate School of Engineering, Fukui University
3-9-1 Bunkyo, Fukui 910-8507, Japan
}%

\begin{abstract}   
Nonlinear coupling between inter- and intra-element dynamics 
appears as a collective behaviour of elements.
The elements in this paper denote 
symptoms such as 
a bacterium having an internal network of genes and proteins, 
a reactive droplet, a neuron in networks, etc.
In order to elucidate the capability of such systems, 
a simple and reasonable model is derived.
This model exhibits the rich patterns of systems such as  
cell membrane, cell fusion, cell growing, cell division, 
firework, branch, and clustered clusters (self-organized hierarchical
 structure, modular network).
This model is extremely simple yet powerful; therefore,  
it is expected to impact several disciplines.
\end{abstract}


\maketitle

We consider non-equilibrium elements whose internal dynamics 
interacts with their macroscopic or mesoscopic order({\it 1-5}). 
An assembly of such elements can describe one aspect of soft-matter; 
there are exciting projects concerning soft matter in several countries including
Japan, where a huge project started recently. 
These reveal the world-wide importance of such studies. 
%
%
In this paper, the elements denote 
symptoms such as 
a bacterium having an internal network of genes and proteins({\it 6}), 
a reactive droplet in reaction-diffusion systems({\it 7}), 
a neuron in networks({\it 8,9}), etc({\it 10-14}).
These elements exhibit 
not only spatio-temporal patterns but also collective functions.
For instance, 
the cohort migration of mammalian cells forms tissue patterns({\it 15,16}), and
the Proteus mirabilis effectively invades human urothelial cells by
swarming({\it 17}).
Further, swarm intelligence has been extensively studied
in order to enable 
a collection of simple robots to perform advanced tasks({\it 18}).
Here, we show a simple model derived by means of mathematical techniques 
to study the cross-cutting phenomenon underlying the above systems 
while ignoring system-specific details. 
The derived model exhibits rich patterns such as 
a modular network and a 
closed membrane that moves around, grows, and multiplies like a cell.  
This model is expected to shed light on issues related to several disciplines.

In order to avoid an ad-hoc toy model,  
we propose a broad class of models from which we derive a simple model
by means of centre-manifold reduction and phase reduction({\it 19}).
At the end of this paper, 
we briefly discuss the vast possibilities of this derived model.  

Self-sustained (or limit-cycle) oscillator is the leading candidate for
the simplest dynamical element. 
Thus, we assume a supercritical Hopf bifurcation for the intra-element dynamics.  
A simple interaction among the elements is mediated by a chemical that
diffuses in space. 
Thus, we assume that the elements exhibit chemotaxis({\it 20}), which means that the elements
are driven by the gradient of chemical density, and the elements produce and
consume the chemical depending on their state.
We carry out centre manifold reduction 
in the neighbourhood of the Hopf bifurcation point.
However, because the reduced model captures the critical centre of systems, 
the model describes the systems in a broad parameter space until another bifurcation occurs.
In addition, the reduced model can be derived 
from another class of models having the same critical centre as that in
our original class of models. 
Thus, the model derived in the following is fairly universal. 

The reduced model of the chemotaxis oscillators is given 
\begin{eqnarray}
\dot{\phi}_i &=& 1+ ( \kappa P({\bm r}_i) + c.c. ), \label{ppp}\\
\dot{\bm{r}}_i &=& - \nabla P(\bm{r})|_{\bm{r}=\bm{r}_i} + c.c. \label{prr}, 
\end{eqnarray}
$\phi_i$ represents the internal state of $i$th element.
The $D$-dimensional real vector $\bm{r}_i$ represents  
the position of the $i$th element. 
$\kappa$ is a complex constant. 
$P$ represents an interaction among the elements and is given by 
\begin{eqnarray}
P(\bm{r}) &\equiv& \sum_j e^{-{\rm
 i}c(\phi_j-\phi_i)}G(\bm{r}_j-\bm{r}), 
\end{eqnarray}
where $c$ is a real parameter. 
Note that the model becomes a potential system when $\psi$ is adiabatically eliminated.
In one-dimensional space,
the coupling kernel $G$ is simply expressed as 
$G^{D=1}(r) = \frac{b}{2\rho} e^{-\rho |r|}$.
$b$ and $\rho$ are complex constants. 
In two-dimensional space,
$G^{D=2}(\bm{r}) = \frac{b}{2\pi} K_0(\rho |\bm{r}|)$,
where $K_0$ is the modified Bessel function of the second kind 
with a complex argument.
In an any-dimensional space, 
$G(\bm{r})$ oscillates and rapidly decreases when $|\bm{r}|$ increases, 
and $G(\bm{r})$ almost vanishes when $|\bm{r}|$ is greater than 
the coupling length $r_c \equiv 1/{\rm Re}\rho$.
Thus, the main characteristic of $G(\bm{r})$ 
- oscillating and decreasing - 
is qualitatively well described by $G^{D=1}(\bm{r})$, 
which we substitute for $G$ in an any-dimensional space 
for simplicity. 
In fact, we numerically confirmed that 
the spatio-temporal patterns shown in this paper can be observed 
for the original $G$ with slight and suitable parameter changes.  

We now rescale \E{ppp} and \E{prr}. 
Introducing a variable $\psi_i$ defined as 
$\psi_i \equiv c \{ \phi_i - [1 + (\kappa G(0) + c.c.) ]t \}$,  
we rescale the spatio-temporal coordinate as $\bm{r'}_i \equiv {\rm Re} \rho \bm{r}_i$ and 
$\partial_{t'} \equiv |\rho/(c \kappa b)| \partial_t$.
Omitting the prime, we obtain
\begin{eqnarray}
\dot{\psi}_i &=& \sum_{j \neq i} e^{-|\bm{R}_{ji}|}
 \sin(\Psi_{ji}+\alpha |\bm{R}_{ji}|-c_1), \label{ppp2}\\
\dot{\bm{r}}_i &=& c_3 \sum_{j \neq i} \hat{\bm{R}}_{ji} e^{-|\bm{R}_{ji}|}
 \sin(\Psi_{ji}+\alpha |\bm{R}_{ji}|-c_2) \label{prr2},
\end{eqnarray}
where $\bm{R}_{ji} \equiv \bm{r}_j - \bm{r}_i$, 
$\hat{\bm{R}}_{ji} \equiv \bm{R}_{ji} / |\bm{R}_{ji}|$  
and $\Psi_{ji} \equiv \psi_j-\psi_i$. 
These equations contain the four real parameters 
$c_1 \equiv \arg(c \kappa b / \rho) - \pi/2$, 
$c_2 \equiv \arg(-b)-\pi/2$,
$c_3 \equiv {\rm Re}\rho |\rho / c \kappa| (>0)$ and 
$\alpha \equiv {\rm Im}\rho / {\rm Re}\rho (>0)$.
Note that $c_3$ is the ratio of the time scales of $\psi_i$ and $\bm{r}_i$. 

Hence, we derived two models: 
model I \E{ppp} and \E{prr}, and 
model II \E{ppp2} and \E{prr2}.
(The second model is equivalent to the first model in one-dimensional
space. 
In a higher-dimensional space, the former is an approximation of the latter
as stated above.) 
These models are extended models reported in the previous paper({\it 21,22}).

Now, we show the richness of these models by using numerical calculations 
carried out in two-dimensional space with a 
periodic boundary condition. 
The boundary conditions are not important in sufficiently
large systems because the coupling function $G(\bm{r})$ decays rapidly 
as $|\bm{r}|$ increases. 
The initial condition is such that 
the positions and phases are randomly distributed.  
The number of elements is fifty.
The figures show snapshots of the element distribution in 
two-dimensional space after the transient time.  
The colours represent the phase $\phi$ (or $\psi$) of the element.
Figure1 shows {\it firework} 
exhibited by the first model. 
This pattern is static in space, 
and the phase waves spread from the centre of the pattern, 
which can correspond to the target pattern in reaction-diffusion systems.
In fact, our models also exhibit spiral waves with another choice of parameters.  
Figure2 shows the {\it closed membrane} 
exhibited by the second model. 
This membrane moves around while maintaining almost its shape.
We can observe the fraction of the membrane where 
the elements are relatively synchronised({\it 23}). 
This fraction is puffs up as time passes.
Thus, in this figure, 
the membrane subsequently moves upwards. 
If we change the parameters slightly, the membrane is 
divided into two synchronous clusters, 
which grow to form two closed membranes, 
and this process repeats.
When the number of elements constituting one membrane decreases 
due to membrane division, the elements merge with neighbouring elements. 
This dynamical pattern reminds us of the proliferation of cells.  
%
Figure3 shows {\it clustered clusters} (or {\it modular networks}) 
exhibited by the second model. 
This self-organised hierarchical structure is constituted
by synchronous clusters that exhibit anti-phase synchronisation with
the neighbouring clusters, 
which reminds us of the self-differentiation of cells.  
Here, it should be noted that in these three patterns, adjacent elements exhibit 
an approximate in-phase synchronisation. 
We cannot present all the patterns in this paper; however,  
by simply changing the parameters, our models can exhibit 
a junction of three branches, 
a crystalline lattice, 
gas, 
collective translational motion parallel to plane phase-wave, 
stick-slip motion of clusters, 
train motion, etc.
\begin{figure}
\resizebox{0.4\textwidth}{!}{\fbox{\includegraphics{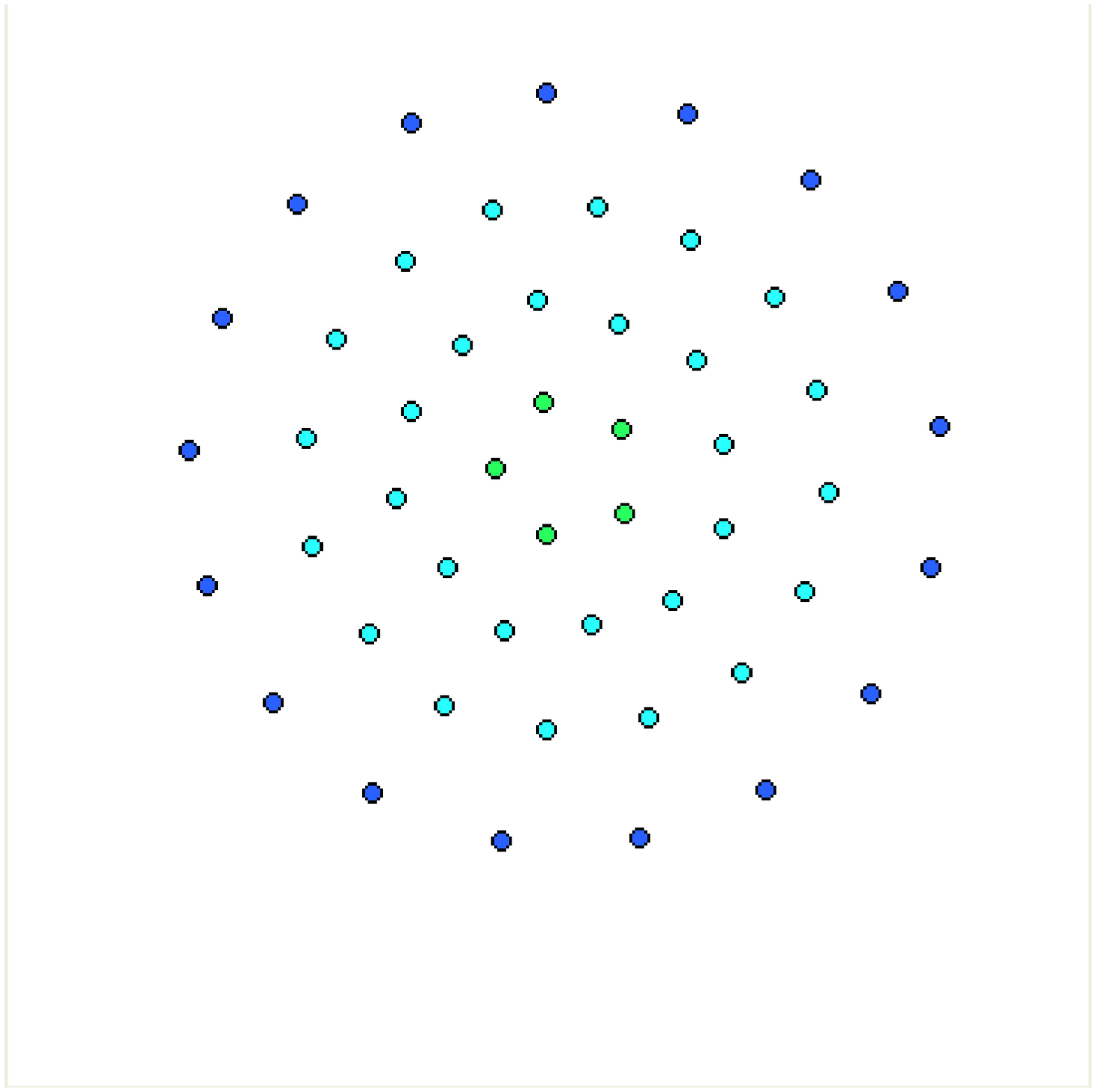}}}
\caption{\label{fig:1}%
{\it Firework}.
Snapshot of element distribution in two-dimensional space.
The colours represent the internal state $\phi$ of the elements. 
The parameters are $\kappa=0.8 + {\rm i} 1$, 
$\rho=0.1$, $b=2 + {\rm i} 3.5$ 
and $c=1$. 
The space size is $120 \times 120$, and it is shown in entirety.
Although we have adopted a point element in this paper, 
we plot its position with a finite size for visualizing.
}
\end{figure}
\begin{figure}
\resizebox{0.4\textwidth}{!}{\fbox{\includegraphics{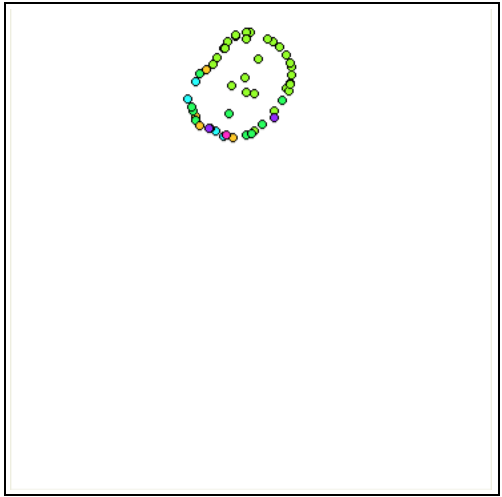}}}
\caption{\label{fig:2}%
{\it Closed membrane} shown in the same manner as Fig.1.
The parameters are $c_1=1.5$, $c_2=3$, $c_3=0.02$ and $\alpha=0$. 
The space size is $10 \times 10$.
}
\end{figure}
\begin{figure}
\resizebox{0.4\textwidth}{!}{\fbox{\includegraphics{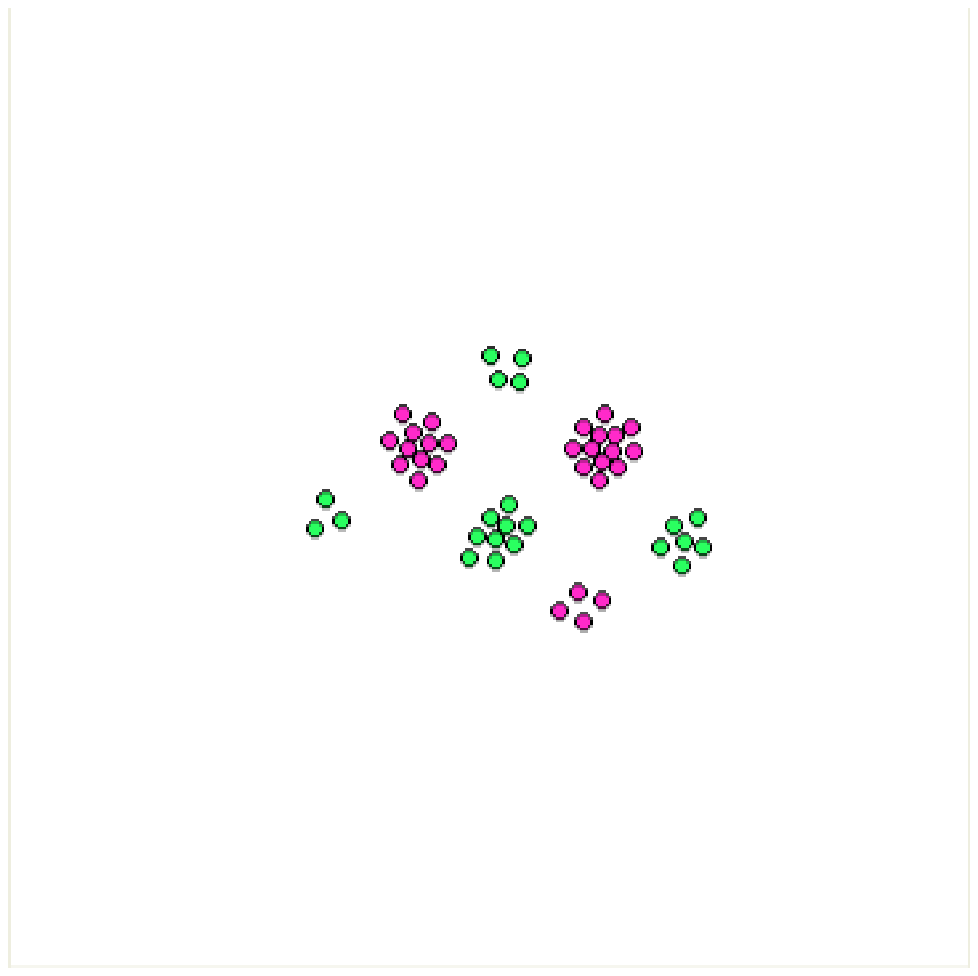}}}
\caption{\label{fig:3}%
{\it Clustered clusters} (or {\it Modular networks}) shown in the same
 manner as Fig.1.
The parameters are $c_1=c_2=c_3=1.5$ and $\alpha=1.6$. 
The space size is $30 \times 30$.
}
\end{figure}

In order to understand the patterns analytically, 
we consider two-oscillators system;  
this corresponds to the case where 
there is only one oscillator in the neighbourhood of the $i$th oscillator. 
This analysis sheds light on many-oscillators system too.
Because $\dot{\bm{r}}_i \| \hat{\bm{R}}_{ji}$, 
the two oscillators move only along a line parallel to 
$\hat{\bm{R}}_{ji}$ that does not change with time. 
Thus, we use $r_i \equiv \bm{r}_i \cdot \hat{\bm{R}}_{ji}$ 
instead of $\bm{r}_i$. 
The difference of the two oscillators 
$\Psi \equiv \psi_2-\psi_1$ and $R \equiv r_2-r_1$
obeys 
$\dot{\Psi} = -2 e^{-R} \cos(\alpha R-c_1) \sin \Psi$, 
$\dot{R} = 2 c_3 e^{-R} \sin (\alpha R-c_2) \cos \Psi$, 
where we assume $R \geq 0$ without losing generality  
because we can transpose the labels of oscillators $1$ and $2$.  
When the first equation is divided by the second equation, 
we can separate variables $\Psi$ and $R$; then,  
integrating once, 
we derive the invariant curve
\begin{eqnarray}
|\sin \Psi| &=& E e^{a_1 R} |\sin(\alpha R-c_2)|^{a_2}, \label{inv}
\end{eqnarray}
where $a_1 \equiv \sin(c_1-c_2)/c_3$ 
and $a_2 \equiv \cos(c_1-c_2)/(\alpha c_3)$.
(We can easily derive another curve when $\alpha=0$ or $c_3=0$.)
$E$ is a conserved quantity that is 
defined by the initial conditions $\Psi(0)$ and $R(0)$.
Thus, the difference of the two oscillators moves on this invariant curve.
This equation implies that if the two oscillators synchronise 
(in-phase $\Psi=0$ or anti-phase $\Psi=\pi$), 
the distance must be $R = c_2/\alpha$ mod ($\pi/\alpha$).
In fact, 
this distance can be observed in Fig.3, where  
the distance between the neighbouring elements in the synchronous cluster is $c_2/\alpha$, 
and the distance between neighbouring synchronous clusters is $c_2/\alpha+\pi/\alpha$.
Further, \E{inv} prohibits 
the distance from making the right hand side of \E{inv} become greater than one.
This implies an effective excluded volume, i.e. 
the elements spontaneously maintain a finite distance between each other 
even if they do not have an excluded volume. 
Another analysis can also be carried out, and we will show this elsewhere.

At the end of this paper, we comment on the vast possibilities of our models.
(1) These models pertain to 
  not only the systems stated in the introductory part   
  but also the following (a)-(f).
 (a) Networks, where the strength of the link (or edge) between nodes corresponds to
     the distance $|\bm{R}_{ij}|$ as shown in \F{5}(a). 
     Actually, {\it closed membrane} and {\it clusterd clusters}
     correspond self-organised small-world network and modular network respectively.
     In a social networking service,  
     the update of an individual page is as frequent as that of the relation
     between individual pages, i.e. 
     the time scale of a node is comparable to that of a link. 
     This is rarely observed in conventional home pages. 
     Our models can shed light on this type of recent network types.
     In addition, our models suggest new possibilities for 
     information or memory processing in neural networks exhibiting
     spike timing dependent synaptic plasticity(STDP)({\it 24}) 
     etc. 
 (b) Fluids where an acoustic wave mediates interaction among the radii of bubbles. 
 (c) Motile-spin glass, where the state of the spin corresponds to
     the phase $\psi$ (or $\phi$), 
     and the wave function of the electron surrounding the spin
     corresponds to the oscillatory coupling function $G$({\it 25}) as
     shown in \F{5}(b).
 (d) Reaction-diffusion systems. 
     For instance, the phase waves on the firework in our models 
     correspond to the target and spiral patterns in reaction-diffusion systems.
     Further, the firework, branch, junction of the three branches and
     cell division in our models can correspond to 
     the spots, stripes, defects in the stripes and pulse division in
     reaction-diffusion systems({\it 26}).
 (e) Frustration systems. 
     In our model with some parameters, the elements stabilize when they are located away
     from each other.  
     This causes frustration in high-density elements systems.
 (f) Time-delayed systems. 
     $\tau \neq 0$ implies an effective delay in the interaction among elements.
(2) Thermodynamical limit, i.e. the presence of many elements, which involves  
    non-equilibrium statistics and an extended kinetic theory of
 gases({\it 27}).
    We can derive a continuous model for the density of elements.  
(3) The coulomb interaction, i.e. galvanotaxis instead of
 chemotaxis({\it 28}).
    This will shed light on ionic fluids where dipoles may correspond to
    the elements.
    Further, this is interesting because of the fact 
    that cancer cells and, more generally, biological cells exhibit galvanotaxis.
(4) Three- or higher-dimensional spaces. We may observe self-organised spherical
     shells. 
(5) Design of materials having newly identified physical
     properties by using the self-organised structures of our
     models({\it 29,30}).

\begin{figure}
\resizebox{0.3\textwidth}{!}{\includegraphics{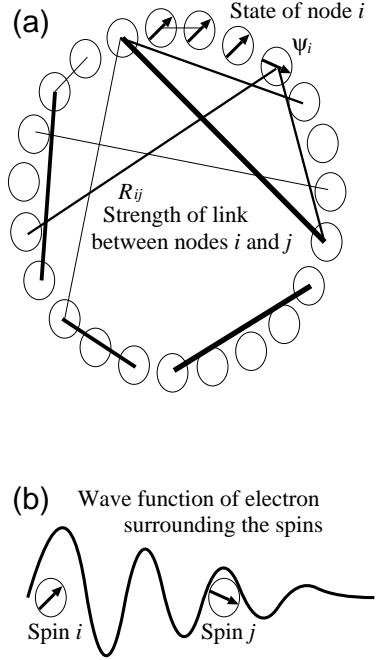}}
\caption{\label{fig:5}%
Our model can be treated as dynamical network (a) or spin glass (b).
}
\end{figure}

There are previous studies (referred in the introductory part)
sharing similar concepts as in this paper. 
However, to date, we have not come across models that are  
as simple, reasonable, powerful and that provide for analytical results. 
Such models are derived in this paper. 
Our models show that only one variable (phase in this paper) 
other than the position variable is sufficient to obtain the rich
collective behaviour of systems.
%

\vspace{1cm}
1.\ 
D.H.Zanette and A.S.Mikhailov, 
{\it Physica D} \textbf{194}, 203-218 (2004).

2.\ 
T.Shibata and K.Kaneko,  
{\it Physica D} \textbf{181}, 197-214 (2003). 

3.\ 
N.Shimoyama, K.Sugawara, T.Mizuguchi, Y.Hayakawa, and M.Sano, 
{\it Phys. Rev. Lett.} \textbf{76}, 3870-3873 (1996). 


4.\ 
S.Sawai, and Y.Aizawa,
{\it J. Phys. Soc. Japan} \textbf{67}, 2557-2560 (1998).

5.\ 
P.Seliger, S.C.Young, and L.S.Tsimring,
{\it Phys. Rev. E} \textbf{65}, 041906 (2002).


6.\ 
T.Zhou, L.Chen, and K.Aihara,
{\it Phys. Rev. Lett.} \textbf{95}, 178103 (2005).

7.\ 
V.K.Vanag, and I.R.Epstein,
{\it Science} \textbf{294}, 835-837 (2001).

8.\ 
D.Watts, and S.Strogatz,
{\it Nature} \textbf{393}, 440-442 (1998).

9.\ 
P.Holme,  B.J.Kim, C.N.Yoon, and S.K.Han,
{\it Phys. Rev. E} \textbf{65}, 056109 (2002).


10.\ 
M.Mimura, M.Nagayama, and T.Ohta,
{\it Methods and Applications of Analysis} \textbf{9}, 493-516 (2002).

11.\ 
P.Jop, Y.Forterre, and O.Pouliquen,
{\it Nature} \textbf{441}, 727-730 (2006).


12.\ 
M.Matsushita, {\it et al.} 
{\it Biofilm} \textbf{1} 305-317 (2004).


%

13.\ 
F.G\"{o}tmark, D.W.Winkler, and M.Andersson,
{\it Nature} \textbf{319}, 589-591 (1986).

14.\ 
S.Dano, P.G.Sorensen, and F.Hynne, 
{\it Nature} \textbf{402}, 320-322 (1999).


15.\ 
G.J.Velicer and Y.T.Yu,
{\it Nature} \textbf{425}, 75-78 (2003).

16.\ 
S.Huang,
C.P.Brangwynne,
K.K.Parker, and
D.E.Ingber,
{\it Cell Motil Cytoskeleton} \textbf{61}, 201-213 (2005).

17.\ 
C.Allison,
N.Coleman,
P.L.Jones, and
C.Hughes,
{\it Infect Immun} \textbf{60}, 4740-4746 (1992).

18.\ 
E.Bonabeau, M.Dorigo, and G.Theraulaz,
{\it Swarm Intelligence: From Natural to Artificial Systems}
(Oxford Univ. Press, New York, USA, 1999).

19.\ 
Y.Kuramoto, {\em Chemical Oscillation, Waves, and Turbulence}
(Springer, New York, USA, 1984); (Dover Edition, 2003).

20.\ 
Y.Miyake, S.Tabata, H.Murakami, M.Yano, and H.Shimizu,
{\it J. theor. Biol.} \textbf{178}, 341-353 (1996).


21.\ 
D.Tanaka and Y.Kuramoto, 
{\it Phys.~Rev.~E} \textbf{68}, 026219 (2003).

22.\ 
D.Tanaka,
{\it Phys.~Rev.~E} \textbf{70}, 015202(R) (2004).

23.\ 
A.Pikovsky, M.Rosenblum, and J.Kurths,
{\it Synchronization: A Universal Concept in Nonlinear Sciences}.
(Cambridge Univ. Press, Cambridge, UK, 2001).

24.\ 
G.Q.Bi and M.M.Poo,
{\it J. Neurosci.} \textbf{15}, 10464-10472 (1998).

25.\ 
J.P.L.Hatchett, I.P.Castillo, A.C.C.Coolen, and N.S.Skantzos,
{\it Phys. Rev. Lett.} \textbf{95}, 117204 (2005).



26.\ 
Y.Nishiura, T.Teramoto, and K.I.Ueda,
{\it Chaos} \textbf{15}, 047509 (2005).

27.\ 
A.Czir\'ok and T.Vicsek, 
{\it Physica A} \textbf{281} 17-29 (2000).

28.\ 
C.E.Pullar {\it et al.}
{\it Mol Biol Cell} \textbf{17}, 4925-4935 (2006).

29.\ 
F.Plenge, H.Varela, and K.Krischer,
{\it Phys. Rev. Lett.} \textbf{94}, 198301 (2005).

30.\ 
S.Sakai, S.Nakanishi, and Y.Nakato,
{\it J. Phys. Chem. B} \textbf{110}, 11944-11949 (2006).

%


\vspace{1cm}
{\bf Acknowledgments}
We thank 
K. Daido, K. Fujimoto, and A. Pikovsky
for valuable comments, 
S. K. Han, M. Nagayama, and H. Nagao for suggetions of possibilities of our models, 
Y. Kawamura, H. Kori, H. Nakao, and H. Kiatahata
for critical reading of the manuscript, 
I. Tsuda, A.S. Mikhailov, K. Showalter, H. Fujisaka, T. Nakagaki
K. Aihara, D. Kurabayashi, A. Ishiguro, A. Awazu, S. Nakata, M. Sano,
K. Sato, T. Shibata, and Y. Kuramoto for encouraging words.
This work was partially supported by a JSPS Research Fellowships for
 Young Scientists and by the Ministry of Education, Science, Sports and
 Culture, Grant-in-Aid for Young Scientists (Start Up), 18840020, 2006.

\end{document}